\documentclass[aps,prb,twocolumn,10pt]{revtex4-2}
\usepackage[utf8]{inputenc}
\usepackage{libertine}
\usepackage[T1]{fontenc}
\usepackage[libertine,upint]{newtxmath} 
\usepackage{graphicx,microtype,mathtools,siunitx,bm,expl3}
\usepackage[cal=boondoxo,frak=boondox]{mathalfa}
\usepackage[colorlinks=true,allcolors=blue]{hyperref}

\DeclareSIUnit\angstrom{\text{Å}}

\DeclarePairedDelimiter\ket{\lvert}{\rangle}

\DeclarePairedDelimiterX\braket[2]{\langle}{\rangle}{#1 \delimsize\vert #2}
\DeclarePairedDelimiterX\matrixel[3]{\langle}{\rangle}{#1 \delimsize\vert #2 \delimsize\vert #3}

\DeclareMathOperator{\eup}{E\!\uparrow}
\DeclareMathOperator{\edn}{E\!\downarrow}
\DeclareMathOperator{\hup}{H\!\uparrow}
\DeclareMathOperator{\hdn}{H\!\downarrow}

\NewDocumentCommand{\grad}{e{_^}}{%
    \mathop{}\!
    \nabla
    \IfValueT{#1}{_{\!#1}}
    \IfValueT{#2}{^{#2}}
}

\makeatletter
\g@addto@macro\@floatboxreset\centering
\makeatother

\begin{document}

\title{Electron pairs bound by the spin-orbit interaction in 2D gated Rashba materials\\ with two-band spectrum}

\author{Yasha Gindikin}
\affiliation{W.I.\ Fine Theoretical Physics Institute, University of Minnesota, Minneapolis, MN 55455, USA}

\author{Igor Rozhansky}
\affiliation{Department of Condensed Matter Physics, Weizmann Institute of Science, Rehovot, 76100, Israel}

\author{Vladimir A.\ Sablikov}
\affiliation{Kotelnikov Institute of Radio Engineering and Electronics, Russian Academy of Sciences, Fryazino branch, Fryazino, 141190, Russia}

\begin{abstract}
We show that the bound electron pairs (BEPs) emerge in two-dimensional gated Rashba materials owing to the interplay of the pair spin-orbit interaction, produced by the Coulomb fields of interacting electrons, and the peculiarities of the band structure giving rise to a negative reduced mass of the interacting electrons. Our consideration is based on the four-band Bernevig-Hughes-Zhang model with the Rashba spin-orbit interaction created by the charges on the gate. The binding energy of the BEP varies with the gate voltage in a wide range across the entire width of the two-particle energy gap. Although the spin-orbit interaction destroys the spin quantization, the BEPs have a magnetic moment, which is created mainly by the orbital motion of the electrons and tuned by the gate voltage.
\end{abstract}
 
\maketitle  
\section{Introduction}

In recent years a great deal of attention is paid to the effects of electron-electron (e-e) interaction in materials with non-trivial band states~\cite{Stemmer_2018}. Of special interest is its interplay with the spin-orbit interaction (SOI)~\cite{Bihlmayer2022,2019arXiv190506340G,maslov_spin,PhysRevB.82.195131,PhysRevB.84.033305}, which under certain circumstances can lead to the purely electronic mechanisms of electron pairing~\cite{Levy_2020}. Studying these mechanisms is important, of course, not only because of problems of high-temperature superconductivity~\cite{kagan2013modern}, but also because exploring the formation of the bound electron pairs (BEPs) and, generally, few-electron complexes~\cite{Kezerashvili2019,https://doi.org/10.1002/qua.25994} lays the ground for understanding a more intricate problem of the correlated many-electron state. The formation of BEPs with the energy lying in the band gap of the many-band material is also interesting because it gives rise to unconventional transport of charge and spin even in the absence of band conductivity.

The BEPs have been studied in graphene and bigraphene~\cite{Sabio2010,MarnhamShytovPRB2015}, in flat-band systems~\cite{Torma,Iskin}, in Dirac semimetals~\cite{Portnoi}, in topological insulators~\cite{PhysRevB.95.085417}, to name just a few. Electron pairing is often attributed to the peculiarities of the band structure and, in particular, to the formation of the negative reduced mass, which governs the relative motion of the electrons in the pair. In this case even the Coulomb repulsive interaction between the electrons can bind them into a pair~\cite{perel1971}.

Another pairing mechanism recently discovered is based on the pair spin-orbit interaction (PSOI). This is the e-e interaction component that depends on the spin and momentum of the electrons, which is produced by the Coulomb fields of interacting electrons in Rashba materials (Ref.~\cite{review_jetp} is a recent review of the subject). The PSOI is attractive for the electrons in particular spin configurations tied to their momentum, and competes with the Coulomb repulsion of electrons. Under certain conditions, attainable in materials with the giant Rashba SOI, the PSOI prevails, which leads to the formation of the BEPs~\cite{PhysRevB.98.115137,10.1016/j.physe.2018.12.028,2019arXiv190409510G}. To date, the BEP formation due to the PSOI was explored within the conduction-band approximation.

This makes it impossible to study the BEPs with high binding energies comparable to the band gap, which is usually small in materials with a very strong SOI\@. However, of greatest interest are the strongly bound pairs considered as composite particles carrying charge and spin. For this reason, the theory of the electron pairing due to the PSOI must be generalized to the multi-band models that properly describe the SOI originating from the hybridization of the atomic orbitals with different spin and orbital structures. This is even more so for the effective reduced mass of an electron pair, the sign and magnitude of which in a multi-band system are determined by the mixture of atomic orbitals involved in the model, which should be found self-consistently from the solution of the equation of motion~\cite{PhysRevB.95.085417}.

The goal of the present paper is to explore the energy spectrum and magnetic properties of the BEPs arising in two-dimensional (2D) systems with the giant Rashba effect. The particular mechanisms behind the giant SOI can be quite different in such systems as the heterostructures based on $\mathrm{LaAlO_3/SrTiO_3}$~\cite{Levy_2015,doi:10.1021/acs.nanolett.8b01614}, the monolayers of $\mathrm{BiSb}$~\cite{PhysRevB.95.165444}, the specifically crafted structures based on 2D layers of van der Waals materials with heavy adatoms~\cite{otrokov2018evidence,PhysRevB.99.085411}, and many others~\cite{Bihlmayer2022}, but in a single-band approximation the main effect of the SOI is captured by a widely used SOI Hamiltonian linear in the wave vector~\cite{bychkov1984properties}. On the contrary, in a multi-band situation the Rashba SOI effect is described by a specific model for each mechanism. 

We base our study on the Bernevig–Hughes–Zhang (BHZ) model~\cite{Bernevig1757}, which is well established to describe many quasi-2D systems in both topological and trivial phases. However, in this model, Rashba SOI is due to the $sp^3$ band hybridization mechanism, which as we found results in a PSOI not strong enough to reliably overcome the Coulomb interaction~\cite{2019arXiv190409510G,GINDIKIN2022115328}. Nonetheless, the situation is significantly improved in the presence of the SOI produced by the electric field of the gate, which makes it easier to achieve this condition.  

Therefore we will study the BEPs in a 2D material described by the BHZ model with the structure-inversion-asymmetry (SIA)  Rashba SOI created by a charged metallic gate. The gate plays several important roles here. The electrons induce the image charges on the gate, the normal electric field of which contributes to the PSOI~\cite{PhysRevB.95.045138,2018arXiv180410826G,2019arXiv190409510G}. The external voltage applied to the gate creates one-particle Rashba SOI that enhances the electron pairing. And finally, the gate screens the Coulomb e-e interaction, which affects the balance between the PSOI and Coulomb interaction by changing the spatial profile of the effective binding potential. 

The SOI is produced by both normal and tangential components of the electric field in the layer. The in-plane electric field is taken by the BHZ model exactly within the $sp^3$ hybridization scheme, whereas the SOI due to the normal electric field is accounted for in the generalization of the original BHZ model~\cite{Rothe_2010}. 

An important point is that the presence of the SOI produced by the electric field due to the gate totally destroys the spin quantization. As a result no spin projection is well defined. The only conserving quantity at a given energy is the normal projection of the total angular moment which is quantized as $J_z = 2m \hbar$, $m \in \mathbb{Z}$. If $J_z$ is nonzero, one can expect the magnetic moment of the BEP to be also nonzero. Considering the BEPs as the composite particles, it is of significant interest to find out the magnitude of the magnetic moment they carry, and how it depends on the energy of the BEP and the system parameters. The predicted non-trivial behavior of the orbital magnetization associated with the BEPs could be probed using the recently emergent state-of-the-art experimental techniques~\cite{Zeldov2022}.

We have shown that the interplay of the PSOI and the peculiarities of the subband structure of the electronic states gives rise to a purely electronic pairing mechanism. The energy level of the resulting BEPs varies with the gate voltage in a wide range across the entire width of the two-particle energy gap. Even though the SOI destroys the spin quantization, BEPs do possess a magnetic moment, which is created mainly by the orbital motion of the electrons and is tunable by the gate voltage.

\section{Model and Results}

The BHZ model is formulated in the four-band basis $\mathcal{B}={(\ket{\eup},\ket{\hup},\ket{\edn},\ket{\hdn})}^{\intercal}$. The states $\ket{\eup}$ and $\ket{\edn}$ are composed of the electron- and light-hole band states with the angular momentum projection of $m_J = \pm 1/2$, whereas $\ket{\hup}$ and $\ket{\hdn}$ are the heavy-hole states with $m_J = \pm 3/2$. The one-body Hamiltonian that governs the envelope wave-function is~\cite{Rothe_2010}:
\begin{equation}
    \label{ham}
    \hat{H} = 
    \begin{bmatrix}
        M + B_1 k^2      & A k_+ & -i \frac{e \xi}{2}\{\mathcal{F}, k_{-}\} &     0\\
        A k_-     & -M -B_2 k^2 & 0          &     0\\
        i \frac{e \xi}{2}\{\mathcal{F}, k_{+}\} & 0     & M + B_1 k^2       & -A k_-\\
        0         &0      & -A k_+     & -M -B_2 k^2
    \end{bmatrix}\,.
\end{equation}
Here $M$ is the mass or gap parameter, $B_1$ and $B_2$ are the dispersion curvatures in the electron-like and hole-like bands, $A$ defines the band hybridization, and $\xi$ is a material-dependent Rashba constant. Then, $k$ is the momentum operator, $k_{\pm} = k_x \pm i k_y$, and curly brackets stand for anti-commutator to ensure the hermiticity for the case of the non-uniform electric field. The normal electric field $\mathcal{F}$ produces the Rashba SOI that couples only the electron-like states. The formation of the bound electron pairs proves highly sensitive to the inversion of the electron-like and hole-like bands. In what follows, we consider only the trivial phase.

In the absence of the SOI, the single-particle spectrum has an energy gap at $(-M,M)$. With Rashba SOI present, the gap extends from the top of the valence band at $\varepsilon = -M$ to the bottom of the conduction band, which is lowered down by the SOI\@. Things change as we go to the two-body problem~\cite{PhysRevB.95.085417}.

\begin{figure}[htb]
    \includegraphics[width=0.9\linewidth]{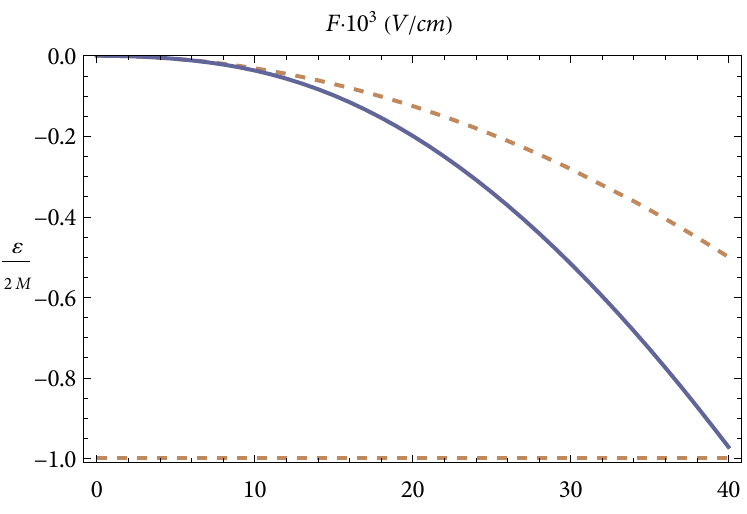}
    \caption{\label{fig1} The energy level of the BEPs (solid line) as a function of the gate electric field $F$ within the energy gap of the continuum of unbound states. 
    The system parameters are $B_1 = \SI{120}{{\angstrom^2} eV}$ and $B_2 = \SI{20}{{\angstrom^2} eV}$. The continuum boundaries are shown by the dashed lines.}
\end{figure}

\begin{figure}[htb]
    \includegraphics[width=0.9\linewidth]{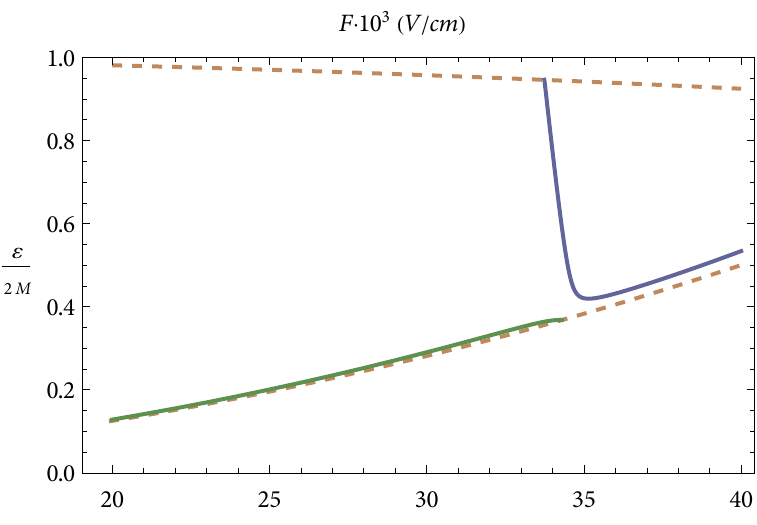}
    \caption{\label{fig2} Two energy levels of the bound electron pairs (green and blue solid lines) as a function of the gate electric field $F$. The continuum boundaries are shown by the dashed lines. The system parameters $B_1 = \SI{20}{{\angstrom^2} eV}$ and $B_2 = \SI{120}{{\angstrom^2} eV}$ are inverted as compared to Fig.~\ref{fig1}.}
\end{figure}

The non-interacting two-body Hamiltonian is constructed as a Kronecker sum $ H_{\mathrm{free}} = H_1 \oplus H_2$ of the one-body terms of Eq.~\eqref{ham}. For electron pairs with zero total momentum, the gap in the two-body spectrum belongs either to the lower (for $B_1 > B_2$) or to the upper (for $B_2 > B_1$) half of the energy interval of $(-2M,2M)$. The position and width of the gap depends on the Rashba SOI magnitude. The continuum boundaries are plotted by the dashed lines for both cases correspondingly in Fig.~\ref{fig1} and Fig.~\ref{fig2}. The energies of the BEPs lie exactly within the gap.

The Hamiltonian of two interacting electrons is obtained by adding the Coulomb interaction energy $\mathcal{U}$ to the free Hamiltonian,
\begin{equation}
    \label{ham2}
    H = H_1 \oplus H_2 + \mathcal{U}(\bm{r}_1-\bm{r}_2) I_{16}\,,
\end{equation}
and by recognizing the fact that the electric field $\bm{\mathcal{E}}(\bm{r})$ of the Coulomb interaction between the electrons depends on their relative in-plane position $\bm{r} \equiv \bm{r}_1 -\bm{r}_2$, which gives rise to the PSOI~\cite{review_jetp}.

The model system we consider consists of a 2D layer of Rashba material proximitized by a charged metallic gate. The gate affects the electron-electron interaction potential
\begin{equation}
    \mathcal{U}(r) = \frac{e^2}{r} -\frac{e^2}{\sqrt{r^2 + a^2}}\,,
\end{equation}
$a/2$ being the distance to the gate. The contribution of the in-plane electric field component $\bm{\mathcal{E}}_{\parallel}(\bm{r}) = \frac{1}{e} \grad_{\bm{r}} \mathcal{U}(r)$ to the PSOI is exactly taken into account within the BHZ model via the hybridization of the electron-like and hole-like bands, similarly to the one-body SOI produced by an external in-plane electric field~\cite{Rothe_2010}. 

The presence of the gate also gives rise to the normal component of the electric field $\mathcal{F}(r) \equiv \mathcal{E}_{\perp}(r) +F$ acting on each of the interacting electrons. It has two terms. The component from the other electron's image
\begin{equation}
    \mathcal{E}_{\perp}(r) = \frac{ea}{{(r^2 +a^2)}^{3/2}}\,,
\end{equation}
which depends on the relative distance between the electrons, contributes to the PSOI~\cite{PhysRevB.95.045138,2019arXiv190409510G}. The uniform component of $e/a^2$ from each electron's own image sums up with the normal field from the net charge of the gate to give the total field $F$, which produces the one-body Rashba SOI\@. 

The two-electron basis is the direct product of $\mathcal{B}_1 \otimes \mathcal{B}_2$. Explicitly, it is written as
\begin{equation}
    \begin{split}
        \mathcal{B}_1 \otimes \mathcal{B}_2 \equiv &(\ket{\eup \eup },\ket{\eup \hup },\ket{\eup \edn },\ket{\eup \hdn },\\
        &\ket{\hup \eup },\ket{\hup \hup },\ket{\hup \edn },\ket{\hup \hdn },\\
        &\ket{\edn \eup },\ket{\edn \hup },\ket{\edn \edn },\ket{\edn \hdn },\\
        &\ket{\hdn \eup },\ket{\hdn \hup },\ket{\hdn \edn },{\ket{\hdn \hdn} )}^{\intercal}\,.
    \end{split}
\end{equation}

\subsection{BEP wave function}
 
For simplicity, we restrict our consideration to a particular case of the electron pairs with zero total momentum, although it is known that the center-of-mass motion of the pair can also lead to electron pairing in Rashba materials~\cite{2018arXiv180410826G,PhysRevB.98.115137}. Such BEPs are formed by two electrons orbiting around their common barycenter.

In this case, the two-electron wave-function, which represents the spinor of the 16th rank, depends only on the relative distance $\mathbf{r} \equiv (r,\varphi)$ between the electrons. The eigenfunctions of the Hamiltonian of Eq.~\eqref{ham2} are also the eigenfunctions of the total angular momentum along the $z$ direction $\hat{J}_z$, which leads to the following form of the wave-function~\cite{PhysRevB.95.085417}:
\begin{equation}
\label{spincom}
    \begin{split}
        \Psi_{m} = e^{2 i m \varphi} &(\psi_{1}(r) e^{-i\varphi},i \psi_{2}(r) e^{-2i\varphi},\psi_{3}(r),i\psi_{4}(r) e^{i\varphi},\\
        &-i\psi_{2}(r) e^{-2i\varphi},\psi_{5}(r) e^{-3i\varphi},i\psi_{6}(r) e^{-i\varphi},\psi_{7}(r),\\
        &-\psi_{3}(r),i\psi_{6}(r) e^{-i\varphi},\psi_{8}(r) e^{i\varphi},-i\psi_{9}(r) e^{2i\varphi},\\
        &i\psi_{4}(r) e^{i\varphi},-\psi_{7}(r),i\psi_{9}(r) e^{2i\varphi},{\psi_{10}(r) e^{3i\varphi} )}^{\intercal}\,,
    \end{split}
\end{equation}
with $m \in \mathbb{Z}$ being the angular quantum number. Note that the spinor contains 10 independent components instead of 16 that would be consistent with the rank of the two-particle Hamiltonian. This is a result of the wave-function anti-symmetry with respect to the electron permutation. The eigenvalue of $J_z$ is $2 m \hbar$. Even values of $2m$ ensure that the wave-function is anti-symmetric under the electron permutation, which is accompanied by $\varphi \to \varphi + \pi$.

The spinor components satisfy the system of coupled equations of Eq.~\eqref{eqmo}. The two-body problem in Rashba materials allows for two seemingly different mechanisms of electron pairing. First, electron pairs may arise bound by the Coulomb interaction because of the negative reduced effective mass of two electrons formed by the hybridization of the electron-like and hole-like bands~\cite{PhysRevB.95.085417,https://doi.org/10.1002/pssr.201900358,https://doi.org/10.1002/pssb.202000299,SABLIKOV2022127956}. On the other hand, the BEPs can be formed due to the PSOI, which proves to be attractive for a particular spin configuration of the electrons. This division is rather arbitrary in the multi-band model, where the BEPs are formed as a result of the interplay of both mechanisms.

The BEPs energy levels are doubly degenerate with respect to the sign of $m$, except for the state with $m=0$, which is non-degenerate. The components of the spinors $\Psi_{m,n_r}$ and $\Psi_{-m,n_r}$, where $n_r \in \mathbb{N}$ stands for the radial quantum number, feature a certain cross-symmetry, which is a direct consequence of the time-reversal invariance in a bound pair composed of electrons with opposite momenta:
\begin{align}
    \label{symm}
    &\Psi_{m,n_r}:        &          &            &\Psi_{-m,n_r}:\notag\\
    &(\ket{\eup \eup })   &\psi_1(r) &= \psi_8(r) &(\ket{\edn \edn })\notag\\
    &(\ket{\eup \hup })   &\psi_2(r) &= \psi_9(r) &(\ket{\hdn \edn })\\
    &(\ket{\eup \hdn })   &\psi_4(r) &= \psi_6(r) &(\ket{\hup \edn })\notag\\
    &(\ket{\hup \hup })   &\psi_5(r) &= \psi_{10}(r) &(\ket{\hdn \hdn })\notag
\end{align}
These equations are also valid for $m=0$. In this case they define the relations between the different components of the spinor $\Psi_{m=0,n_r}$.

\subsection{BEP spectrum}

The eigenvalue problem of Eq.~\eqref{eqmo} is treated numerically. The system spectrum is found by the finite elements method~\cite{zienkiewicz} using the Arnoldi eigenvalue solver~\cite{golubvanloan}. The method of choice to find all the eigenvalues in the energy gap would be the FEAST solver~\cite{FEAST}, were it not for the lack of support for non-hermitian problems in its early version currently used in the Intel Math Kernel Library. 

We choose the system parameters typically found in 2D Rashba materials~\cite{manchon2015new}:  $ M = \SI{0.01}{eV}$, $A = \SI{5}{\angstrom eV}$, $\epsilon = 20$. However, we assume larger SOI coupling $\xi =\SI{5 e3}{{\angstrom^2}}$, as contrasted to $\xi$ from $\SI{1 e3}{{\angstrom^2}}$ to $\SI{2 e3}{{\angstrom^2}}$ commonly found in such system, and quite a short distance of $a = \SI{20}{\angstrom}$. 

Figure~\ref{fig1} shows the position of the energy level of a BEP as a function of the gate electric field $F$. This is the BEP ground state corresponding to $m=0$ and radial quantum number of $n_r=1$. The one-body Rashba SOI produced by the electric field $F$ is seen to enhance the electron pairing by increasing the binding energy, in agreement with a conduction-band approximation treatment~\cite{2018arXiv180410826G,2019arXiv190409510G}. The BEP in the ground state has no definite spin, with zero spin projection along any direction. There also exist shallow energy levels close to the lower continuum boundary, not resolvable in the figure.

\subsection{BEP helical structure and magnetism}

The fundamental properties of the BEPs as composite particles that can manifest themselves in a collective behavior of the many-body system are their charge and spin. While the charge is well known, it obviously equals $2e$, the spin is not defined. However, a composite particle certainly has magnetic properties, which are determined by the helical structure of its quantum state, which is formed by the relative motion of paired electrons and the contributions of different atomic orbitals to the spinor of the state.

Before presenting the results, let us dwell in more detail on the essence of the problem. First we note that the helicity of single-particle states, which physically means the tight coupling of the momentum to the spin, arises due to the strong SOI\@. In the model under study the SOI is inherently presented through the $sp^3$ hybridization. Additional factors producing the SOI, such as a natural inversion asymmetry caused by the atomic structure of the 2D layer or the electric fields of external charges, break down the spin quantization so that no spin projection is defined and therefore the helicity can not be determined. This, however, does not mean that the helical nature of the quantum state is destroyed~\cite{PhysRevLett.108.156402}. The point is that the system remains symmetrical with respect to the time reversal and therefore the eigenstates are classified by their Kramers index instead of the spin projection. In each of the Kramers partners the electrons move in opposite direction and their spin states are described by different four-component spinors related by the time reversal transformation~\cite{PhysRevB.102.075434}. The two-particle bound states forming under this condition have been extremely poorly studied. Since the spin is not a quantity that characterizes these states, and only the total angular momentum is a conserved quantity, the magnetic moment of a pair is the measurable quantity that most adequately characterizes its helical structure. 

The total angular momentum $\hat{J}_z = \hat{L}_z + \hat{S}_z +\hat{K}_z$ of the BEP has three components: the orbital moment $\hat{L}_z$ of the relative motion of the particles, the electron spin $\hat{S}_z$, and the angular momentum $\hat{K}_z$ of the atomic orbitals. All three contribute to the magnetic moment $\mu = \mu_{\mathrm{orbital}} +  \mu_{\mathrm{spin}} +  g \mu_{\mathrm{Bloch}}$  of the BEP, with material-dependent Landé factor $g$. 

The orbital magnetization
\begin{widetext}
\begin{equation}
    \begin{split}
        \frac{\mu_{\mathrm{orbital}}}{\mu_B} = &{} \frac{m}{\hbar^2}\int_0^{\infty} 
        \Big[-B_1 \left((2m-1)(\psi_1^2(r) +\psi_6^2(r)) + (2m-2)\psi_2^2(r) +4m\psi_3^2(r) +(2m+1)(\psi_4^2(r) +\psi_8^2(r))  +(2m+2)\psi_9^2(r)\right) \\
        &{} +B_2 \left((2m-2)\psi_2^2(r) + (2m+1)\psi_4^2(r) + (2m-3)\psi_5^2(r) +(2m-1)\psi_6^2(r) +4m\psi_7^2(r)  +(2m+2)\psi_9^2(r) +(2m+3)\psi_{10}^2(r)\right) \\
        &{} - r A \left( \psi_2(r)(\psi_1(r)+\psi_5(r)) +(\psi_4(r) -\psi_6(r))(\psi_3(r)+\psi_7(r)) -\psi_9(r)(\psi_8(r)+\psi_{10}(r))\right)\\
        &{} - e \xi r \mathcal{F}(r) \left(\psi_1(r)\psi_3(r) -\psi_2(r)\psi_6(r) -\psi_3(r)\psi_8(r) +\psi_4(r)\psi_9(r)\right) \Big] 4 \pi r \, dr 
    \end{split}
\end{equation}
\end{widetext}
is produced by the circular current due to the relative motion of the electrons within the pair. The magnetic moment is strongly renormalized by the SOI, which affects the electron velocity.

The spin magnetization
\begin{equation}
    \begin{split}
        \frac{\mu_{\mathrm{spin}}}{\mu_B} = {}& \int 
         \big(-\psi_1^2(r) -2\psi_2^2(r) -\psi_5^2(r)\\
         &{} +\psi_8^2(r) +2\psi_9^2(r) +\psi_{10}^2(r)\big)4 \pi r \, dr 
    \end{split}
\end{equation}
is due to the Zeeman splitting of the spin sector. The last contribution to the magnetization
\begin{equation}
    \begin{split}
        \frac{\mu_{\mathrm{Bloch}}}{\mu_B} = {}& \int 
         \big(-\psi_2^2(r) +\psi_4^2(r) -\psi_5^2(r)\\
         &{} -\psi_6^2(r) +\psi_9^2(r) +\psi_{10}^2(r)\big) 4 \pi r \, dr 
    \end{split}
\end{equation}
is due to the angular moment of the $p$-like basis Bloch states. 

The lowest energy level ($n_r=1,m=0$) possesses no magnetic moment associated with any of them, as all three contributions vanish by force of Eq.~\eqref{symm}. However, the excited states carry the circular current hence contributing to the orbital magnetization, possess the non-zero spin projection on the normal direction, and also feature the finite magnetization coming from the atomic orbitals.

The formation of a BEP with a nonzero angular momentum due to PSOI encounters certain limitations caused by the presence of a centrifugal barrier, which rapidly grows with decreasing the distance from the center and prevents the attraction due to the PSOI\@. Calculations show that BEPs with nonzero angular momentum are formed at a sufficiently small dispersion parameter $B_1$ in the electron subband. This is not difficult to understand, since within the framework of the used model the SOI acts on the electron subband states, therefore an increase in the effective mass leads to a decrease in the centrifugal energy, and the attraction due to the PSOI becomes predominant. Further study of the magnetic moment is carried out for this situation, which is quite realizable due to the advances in the band engineering~\cite{https://doi.org/10.1002/smll.201402380}.

\begin{figure}[htb]
    \includegraphics[width=0.9\linewidth]{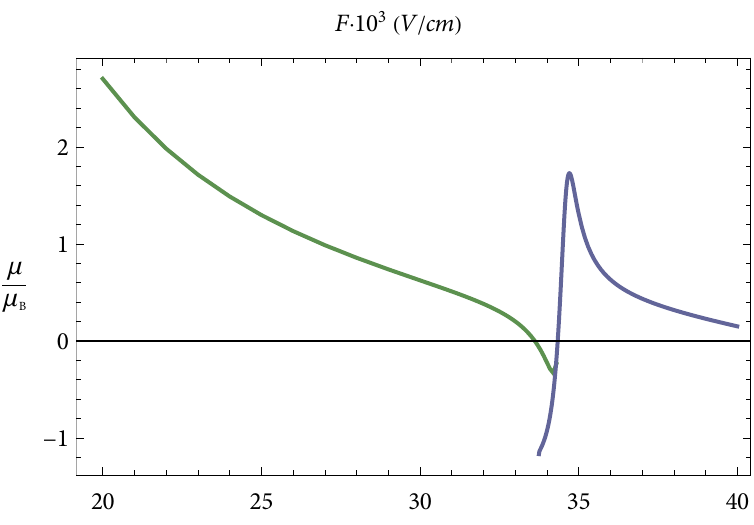}
    \caption{\label{fig3} The orbital magnetization of both types of the BEPs as a function of the gate electrical field.}
\end{figure} 
 
The energy spectrum of the BEPs with $n_r=1,m=1$ for this case is shown in Fig.~\ref{fig2} as a function of the gate voltage. Two different types of the bound states are seen to exist. At low gate voltage, the BEP is formed with the energy slightly above the continuum boundary. As the gate electric field increases, the energy of the state increases too. The second type of the BEP arises for the electrical field larger the some critical value $F_c$, which is determined by a balance between the  Coulomb repulsion of the electrons, the sign and magnitude of their effective reduced mass, and the effect of the PSOI\@. Its energy features the non-monotonous dependence on the gate field, which strongly affects the hybridization of the electron bands that form the BEP\@. When its energy becomes comparable with that of the shallow state, a visible anti-crossing appears in the spectrum.

The orbital magnetic moment $\mu_{\mathrm{orbital}}$ of two types of the BEPs is plotted versus $F$ in Fig.~\ref{fig3}. Both the magnitude and sign of the $\mu_{\mathrm{orbital}}$ are tunable by the gate electric field. Increasing $F$ leads to a strong reconstruction of the orbital and even helical composition of the BEPs.

\begin{figure}[htb]
    \includegraphics[width=0.9\linewidth]{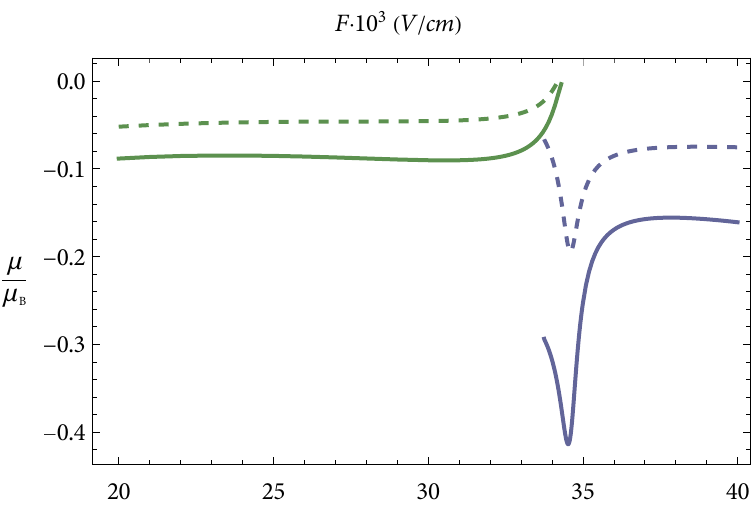}
    \caption{\label{fig4} The spin magnetization (solid lines) and magnetization due to the atomic orbitals (dashed lines) of both types of the BEPs as a function of the gate electrical field.}
\end{figure}

The BEP magnetization is dominated by the orbital contribution, as can be seen from the comparison with Fig.~\ref{fig4}, where the spin magnetization and the magnetization due to the basis atomic orbitals are plotted as a function of the gate electric field.

\section{Conclusion}

We have studied the formation of the BEPs due to the PSOI in a gated 2D material with a two-band spectrum and a strong Rashba SOI created due to the $sp^3$ band hybridization. The gate plays an important role here as it creates both the electric field normal to the layer, which is homogeneous along the plane, and the electric field of the image charges of the electrons, which contributes to the PSOI along with the in-plane component of the Coulomb field of interacting electrons. Due to the presence of the gate, the influence of the PSOI on the BEPs emerging in the system increases significantly, and can be controlled by the gate field. Thus, it becomes possible to shed light on the physical mechanism of the formation of the BEPs. 

Two mechanisms are discussed that give rise to the effective attraction of the electrons: the universal mechanism for the formation of a negative effective reduced mass due to the fact that the quantum state of a pair is formed by mixing the subband states with opposite masses, and the mechanism of the PSOI, created by the Coulomb fields of electrons in Rashba materials. Both are essentially two aspects of the subband hybridization, but reflect different properties of the subband states. By changing the gate field, one can trace how the interplay of these two mechanisms manifests itself in the spectrum and magnetic properties of the BEPs.

We found that an increase in the PSOI leads to the appearance of the BEPs with a sufficiently high binding energy, the energy level of which, with an increase in the gate field, moves from the top of the gap in the two-particle spectrum almost to its bottom. Along with these pairs, there are the states with an energy slightly above the bottom of the gap, which also exist without a SOI\@. Under certain conditions the levels of these two states can converge and, as a result of the avoided crossing, the states hybridize.

The gate-induced SOI breaks the spin quantization, because of which the spin projection in any direction is no longer defined. The quantum state of the BEP is characterized only by the normal projection of the total angular momentum. Nevertheless, the composite particle formed as a result of the electron pairing has a magnetic moment directed normally to the layer. Its magnitude is determined mainly by the orbital relative motion of the electrons, although the motion of the spin, and the magnetic moment of the atomic orbitals also contribute to it. In the ground state with zero angular momentum the magnetic moment is also zero. The magnetic moment exists in the states with nonzero total angular momentum. The magnitude of the magnetic moment can be fine-tuned by the gate field, giving rise to a novel mechanism of the electrically controlled orbital magnetization of the two-electron bound states.

The dependence of the magnetic moment on the gate field is quite different for the pairs formed mainly due to the negative effective reduced mass or mainly due to the PSOI\@. The magnetic moment can even change its sign in the course of the hybridization of the states, which indicates a rearrangement of the helical structure of the BEPs. This apparently stems from a change in the relative motion of electrons during the hybridization of the upper and lower branches of the spectrum of the BEPs with predominantly different pairing mechanisms. Such a sign change would be an experimental fingerprint of the discussed mechanism.
 
\begin{acknowledgments}
    Y.G.\ and I.R.\ gratefully acknowledge the support and hospitality of the Weizmann Institute of Science during their stay there. The work of V.A.S.\ was carried out in the framework of the state task for the Kotelnikov Institute of Radio Engineering and Electronics and partially supported by the Russian Foundation for Basic Research, Project No.~20--02--00126.
\end{acknowledgments}

\appendix
\section{Equation of motion}
Hamiltonian of Eq.~\eqref{ham2} leads to the following system of coupled equation of motion for the spinor components of Eq.~\eqref{spincom}:
\begin{widetext}
    \begin{equation}
        \label{eqmo}
        \begin{split}
            [-2B_1(\partial_r^2 +\tfrac{1}{r} \partial_r -\tfrac{{(2m-1)}^2}{r^2})+2M +\mathcal{U}(r)]\psi_1(r) -2A(\partial_r -\tfrac{2m-2}{r})\psi_2(r) +2 e \xi \mathcal{F}(r)(\partial_r+\tfrac{2m}{r}) \psi_3(r) + \psi_3(r)e \xi \partial_r \mathcal{F}(r) &= \varepsilon \psi_1(r) \\
            A(\partial_r +\tfrac{2m-1}{r})\psi_1(r) + [(B_2-B_1)(\partial_r^2 +\tfrac{1}{r} \partial_r -\tfrac{{(2m-2)}^2}{r^2}) +\mathcal{U}(r)]\psi_2(r)-A(\partial_r -\tfrac{2m-3}{r})\psi_5(r)& \\
            {} -e \xi \mathcal{F}(r)\partial_r\psi_6(r) - e \xi (\tfrac{2m-1}{r}\mathcal{F}(r) +\tfrac12 \partial_r \mathcal{F}(r))\psi_6(r)  &= \varepsilon \psi_2(r) \\
                -e \xi \mathcal{F}(r)\partial_r\psi_1(r) + e \xi (\tfrac{2m-1}{r}\mathcal{F}(r) -\tfrac12 \partial_r \mathcal{F}(r))\psi_1(r)+[-2B_1(\partial_r^2 +\tfrac{1}{r} \partial_r-\tfrac{{(2m)}^2}{r^2} )+2M +\mathcal{U}(r)]\psi_3(r)&\\
                + A(\partial_r +\tfrac{2m+1}{r})\psi_4(r)
                + A(\partial_r -\tfrac{2m-1}{r})\psi_6(r) - e \xi \mathcal{F}(r)\partial_r\psi_8(r) - e \xi (\tfrac{2m+1}{r}\mathcal{F}(r) +\tfrac12 \partial_r \mathcal{F}(r))\psi_8(r) &= \varepsilon \psi_3(r)\\
                -A (\partial_r -\frac{2m}{r}) \psi_3(r) + [(B_2-B_1)(\partial_r^2 +\tfrac{1}{r} \partial_r -\tfrac{{(2m+1)}^2}{r^2}) +\mathcal{U}(r)]\psi_4(r) -A (\partial_r -\frac{2m}{r}) \psi_7(r)& \\
                {} + e \xi \mathcal{F}(r)\partial_r\psi_9(r) + e \xi (\tfrac{2m+2}{r}\mathcal{F}(r) +\tfrac12 \partial_r \mathcal{F}(r))\psi_9(r) &= \varepsilon \psi_4(r)  \\
            2A(\partial_r +\tfrac{2m-2}{r})\psi_2(r) + [2B_2(\partial_r^2 +\tfrac{1}{r} \partial_r -\tfrac{{(2m-3)}^2}{r^2})-2M +\mathcal{U}(r)]\psi_5(r) &= \varepsilon \psi_5(r) \\
            e \xi \mathcal{F}(r)\partial_r\psi_2(r) - e \xi (\tfrac{2m-2}{r}\mathcal{F}(r) -\tfrac12 \partial_r \mathcal{F}(r))\psi_2(r) -A (\partial_r +\frac{2m}{r}) \psi_3(r)&\\
             + [(B_2-B_1)(\partial_r^2 +\tfrac{1}{r} \partial_r -\tfrac{{(2m-1)}^2}{r^2}) +\mathcal{U}(r)]\psi_6(r)
            -A (\partial_r +\frac{2m}{r}) \psi_7(r) &= \varepsilon \psi_6(r)  \\
            A(\partial_r +\tfrac{2m+1}{r})\psi_4(r) +A(\partial_r -\tfrac{2m-1}{r})\psi_6(r) + [2B_2(\partial_r^2 +\tfrac{1}{r} \partial_r -\tfrac{{(2m)}^2}{r^2})-2M +\mathcal{U}(r)]\psi_7(r) &= \varepsilon \psi_7(r) \\
            2 e \xi \mathcal{F}(r)(\partial_r-\tfrac{2m}{r}) \psi_3(r) + \psi_3(r) e \xi \partial_r \mathcal{F}(r) +[-2B_1(\partial_r^2 +\tfrac{1}{r} \partial_r -\tfrac{{(2m+1)}^2}{r^2})+2M +\mathcal{U}(r)]\psi_8(r) -2A(\partial_r +\tfrac{2m+2}{r})\psi_9(r)  &= \varepsilon \psi_8(r) \\
            - e \xi \mathcal{F}(r)\partial_r\psi_4(r) + e \xi (\tfrac{2m+1}{r}\mathcal{F}(r) -\tfrac12 \partial_r \mathcal{F}(r))\psi_4(r) +A(\partial_r -\tfrac{2m+1}{r})\psi_8(r) &\\
            {}+ [(B_2-B_1)(\partial_r^2 +\tfrac{1}{r} \partial_r -\tfrac{{(2m+2)}^2}{r^2}) +\mathcal{U}(r)]\psi_9(r)
             -A(\partial_r +\tfrac{2m+3}{r})\psi_{10}(r) &= \varepsilon \psi_9(r) \\
            2A(\partial_r -\tfrac{2m+2}{r})\psi_9(r) + [2B_2(\partial_r^2 +\tfrac{1}{r} \partial_r -\tfrac{{(2m+3)}^2}{r^2})-2M +\mathcal{U}(r)]\psi_{10}(r) &= \varepsilon \psi_{10}(r)\,.
        \end{split}
    \end{equation}
\end{widetext}

\bibliography{paper}

\end{document}